\documentclass[runningheads]{llncs}
\usepackage[bookmarksopen,
bookmarksopenlevel=1,
bookmarksdepth=2,
pdftex,
pdfauthor={Tim Kräuter},
pdftitle={Blazingly Fast, Comprehensible, and Fixable Soundness Checking of Realistic BPMN Models},
pdfsubject={BPMN soundness checking},
pdfkeywords={BPM, Verification, Soundness, Safeness}
]{hyperref}
\usepackage{graphicx}
\usepackage{tabularray}
\usepackage{listings, listings-rust}
\usepackage{orcidlink} % Orcid links
\newcommand{\subpart}[1]{\vspace{1em}\noindent\textbf{#1}}

\begin{document}
\title{Instantaneous, Comprehensible, and Fixable Soundness Checking of Realistic BPMN Models}
\titlerunning{Instantaneous, Comprehensible, and Fixable BPMN Soundness Checking}
\author{
	Tim Kr\"{a}uter\inst{1}\orcidlink{0000-0003-1795-0611} \and
	Patrick St\"{u}nkel\inst{1}\orcidlink{0000-0002-0537-295X} \and
	Adrian Rutle\inst{1}\orcidlink{0000-0002-4158-1644} \and
	Harald K\"{o}nig\inst{2,1}\orcidlink{0000-0001-6304-6311} \and
	Yngve Lamo\inst{1}\orcidlink{0000-0001-9196-1779}
}
\authorrunning{T. Kräuter et al.}
\institute{Western Norway University of Applied Sciences, Bergen, Norway \\
\email{tkra@hvl.no, patrick.stuenkel@hvl.no, aru@hvl.no, yla@hvl.no} \and
University of Applied Sciences, FHDW, Hanover, Germany\\
\email{harald.koenig@fhdw.de}}

\maketitle
\begin{abstract}
Many business process models have control-flow errors, such as deadlocks, which can hinder proper execution.
In this paper, we introduce our new soundness-checking tool that can instantaneously identify errors in BPMN models, make them comprehensible for modelers, and even suggest corrections to resolve them automatically.
We demonstrate that our tool's soundness checking is instantaneous, i.e., it takes less than 500ms, by benchmarking our tool against synthetic BPMN models with increasing size and state space complexity, as well as realistic models provided in the literature.
Moreover, the tool directly displays possible soundness violations in the model and provides an interactive counterexample visualization of each violation.
Additionally, it provides fixes to resolve the violations found, which are not currently available in other tools.
The tool is open-source, modular, extensible, and integrated into a popular BPMN modeling tool.

\end{abstract}

% Redefine enumerations
\renewcommand{\labelenumi}{(\textbf{\arabic{enumi})}}

\section{Introduction} \label{sec:introduction}

% Problem statement
Business Process Modeling Notation (BPMN) is becoming increasingly popular for automating processes and orchestrating people and systems.
However, many process models suffer from control-flow errors, such as deadlocks and lack of synchronization~\cite{fahlandAnalysisDemandInstantaneous2011}.
These errors hinder the correct execution of BPMN models and may be detected late in the development process, resulting in elevated costs.

\begin{figure}[ht]
	\centering
	\includegraphics[width=0.6\textwidth]{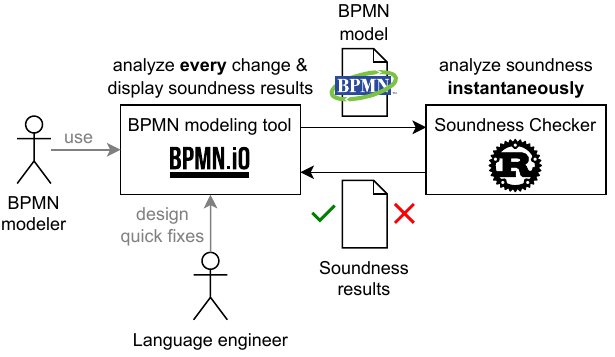}
	\caption{Overview of the tool}
	\label{fig:overview}
\end{figure}

% Solution
In this paper, we describe our new tool, which checks BPMN process models for soundness and safeness~\cite{corradiniClassificationBPMNCollaborations2018}, which entails finding control-flow errors already during modeling.
\autoref{fig:overview} shows an overview of our tool, which we provide online~\cite{timkrauterBPM2024Artifacts2024}.
The tool front-end is based on the popular \textit{bpmn.io} ecosystem, while the soundness checker is implemented in Rust for fast performance, memory efficiency, and memory safety.
% Even blazingly fast performance

The tool can check models after each change since soundness checking is \textit{instantaneous} according to \cite{fahlandAnalysisDemandInstantaneous2011}, i.e., it takes 500ms or less.
Furthermore, we ensure the results are \textit{comprehensible} by highlighting possible violations directly in the model and displaying an interactive counterexample visualization.
Finally, the tool suggests \textit{fixes} for the most common soundness violations and can be extended to suggest more fixes independent of the soundness checker.
However, the tool is still a \textit{prototype} and cannot provide fixes for all possible violations due to the wide variety of control-flow errors.

% Paper structure
In the remainder of the paper, we describe how instantaneous, comprehensible, and fixable soundness checking is achieved in \autoref{sec:soundness}.
Then, we explain our tool implementation in \autoref{sec:impl} before presenting related work (\autoref{sec:related-work}).
Finally, we discuss possible limitations in \autoref{sec:limitations} and conclude in \autoref{sec:conclusion}.

\section{Soundness Checking of BPMN Models} \label{sec:soundness}

% Describe soundness checking for BPMN models.
Soundness was introduced for workflow nets in~\cite{vanderaalstApplicationPetriNets1998} and stems from the field of Petri Nets.
We will use the formal definition directly given for BPMN by~\cite{corradiniClassificationBPMNCollaborations2018}.
Soundness is composed of the three following sub-properties~\cite{corradiniClassificationBPMNCollaborations2018}:
\textit{(i) Option to complete}: any running process instance must eventually complete,
\textit{(ii) Proper Completion}: at the moment of completion, each token of the process instance must be in a different end event and \textit{(iii) No dead activities}: any activity can be executed in at least one process instance.
Option to complete is vital since it guarantees the absence of deadlocks, while proper completion is less important, but it helps enforce BPMN best practices, such as not reusing end events.
Furthermore, It is crucial not to have dead activities since, similar to dead code, they should be removed or they indicate a deeper issue that must be investigated.

In addition, we check \textit{Safeness} to find possible missing synchronizations.
A BPMN model is \textit{safe} if, during its execution, no more than one token occurs simultaneously along the same sequence flow~\cite{corradiniClassificationBPMNCollaborations2018}.
Safeness helps to find \textit{lack of synchronization}~\cite{fahlandAnalysisDemandInstantaneous2011}, for example, if outgoing sequence flows of a parallel gateway are merged using an exclusive gateway.
This would lead to all subsequent activities being executed twice, which is often not desired.

In the remainder of this section, we describe how the tool achieves \textit{instantaneous}, \textit{comprehensible}, and \textit{fixable} soundness checking of realistic BPMN models.

\subsection{Instantaneous Soundness Checking} \label{subsec:instantaneous}
\textit{Instantaneous} soundness checking is defined as taking 500 ms or less in~\cite{fahlandAnalysisDemandInstantaneous2011}.
In this section, we demonstrate that our soundness checker is instantaneous by validating it from three viewpoints.
First, we investigate how our tools react to rapidly \textit{growing model size}.
We use a benchmark based on synthetically generated BPMN models.
Second, we study how our tool deals with a \textit{growing number of parallel branches} of varying size.
Third, we run soundness checking for realistic BPMN process models available in the literature and public datasets.

% General info
For all our benchmarks, we use the hyperfine benchmarking tool~\cite{peterHyperfine2023} (version 1.18.0), which calculates the average runtime when executing each soundness check ten or more times.
The benchmarks were run on Ubuntu 22.04.4 with an AMD Ryzen 7700X processor (4.5GHz) and 32 GB of RAM (5600 MHz).
All used BPMN models, our tools to generate them, and benchmarking scripts are given in the artifacts of this paper~\cite{timkrauterBPM2024Artifacts2024}.

\begin{figure}[ht]
	\centering
	\includegraphics[width=0.8\textwidth]{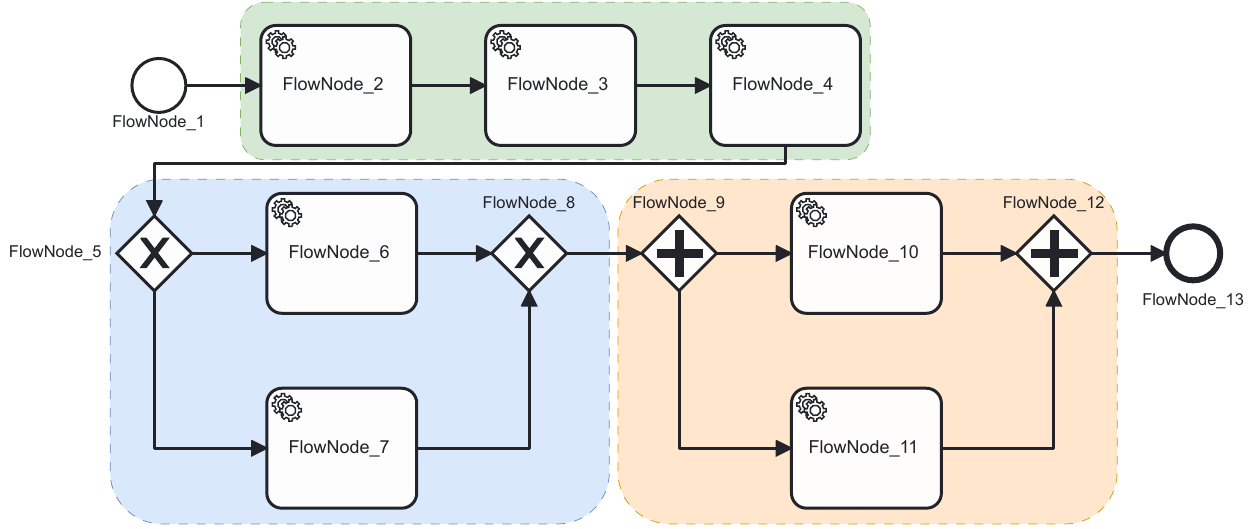}
	\caption{BPMN model with three blocks~\cite{krauterHigherorderTransformationApproach2024}}
	\label{fig:three-block-example}
\end{figure}

\subpart{Growing Model Size}
We use an extended version of the data set of models provided in~\cite{krauterHigherorderTransformationApproach2024}, which consists of 300 synthetically generated BPMN models, and increase it to 500 models.
% Briefly recap the method.
Every BPMN model contains a start event, a fixed amount of \textit{blocks}, and an end event.
The three unique blocks are shown in \autoref{fig:three-block-example}.
To generate the models, we start with the first block and add one more block at a time, always following the same order until we reach 500.
\autoref{fig:three-block-example} shows the third BPMN model in the data set, which contains each block once~\cite{krauterHigherorderTransformationApproach2024}.

\begin{figure}[ht]
	\centering
	\includegraphics[width=0.6\textwidth]{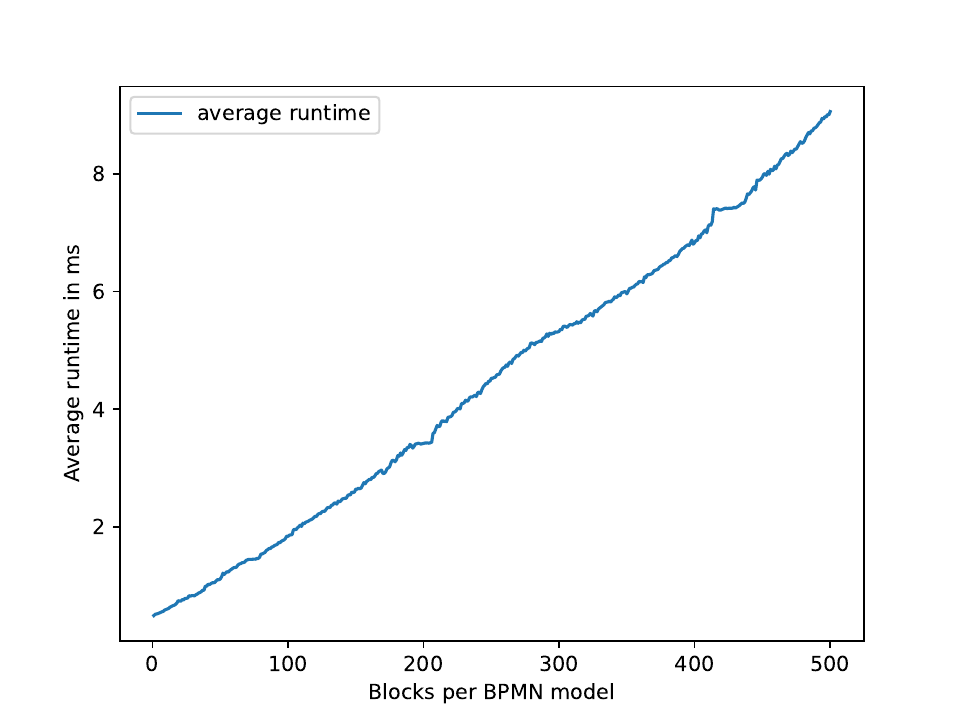}
	\caption{Soundness checking runtime for models of increasing size}
	\label{fig:model-size-benchmark}
\end{figure}

\autoref{fig:model-size-benchmark} shows the average runtime of our tool when checking soundness and safeness for the BPMN models with increasing model size.
Our tool explores the entire state space while simultaneously verifying all properties.
\autoref{fig:model-size-benchmark} shows a linear increase in runtime since the state space increases linearly.

% Results
Our tool spends from 1 ms up to 9 ms for the BPMN models compared to 0.7 s up to 14 s in~\cite{krauterHigherorderTransformationApproach2024} for only 300 blocks on a similar machine.
Thus, our tool can be classified as \textit{instantaneous} according to~\cite{fahlandAnalysisDemandInstantaneous2011} even if the model size increases to over 4000 BPMN elements (500 blocks, 2168 states).
Models of this size are not human-readable anymore and are usually divided into smaller models according to best practices~\cite{fahlandAnalysisDemandInstantaneous2011}.
Thus, models found in practice are likely to be smaller than those found in our benchmark.
However, they might be more complex, leading to a larger state space, as discussed in the next section.

\subpart{Growing Number of Parallel Branches}
An increased model size leads to a bigger state space that must be analyzed.
In the previous section, a linear increase in model size led to a similar growth in the state space.
However, models with a growing number of parallel branches lead to an exponential increase in the state space, i.e., a state space explosion~\cite{clarkeHandbookModelChecking2018}.
In this section, we benchmark our tool against a synthetic data set of models that lead to a state space explosion.
This represents a \textit{worst case} scenario for soundness checking.

\begin{figure}[ht]
	\centering
	\includegraphics[width=0.7\textwidth]{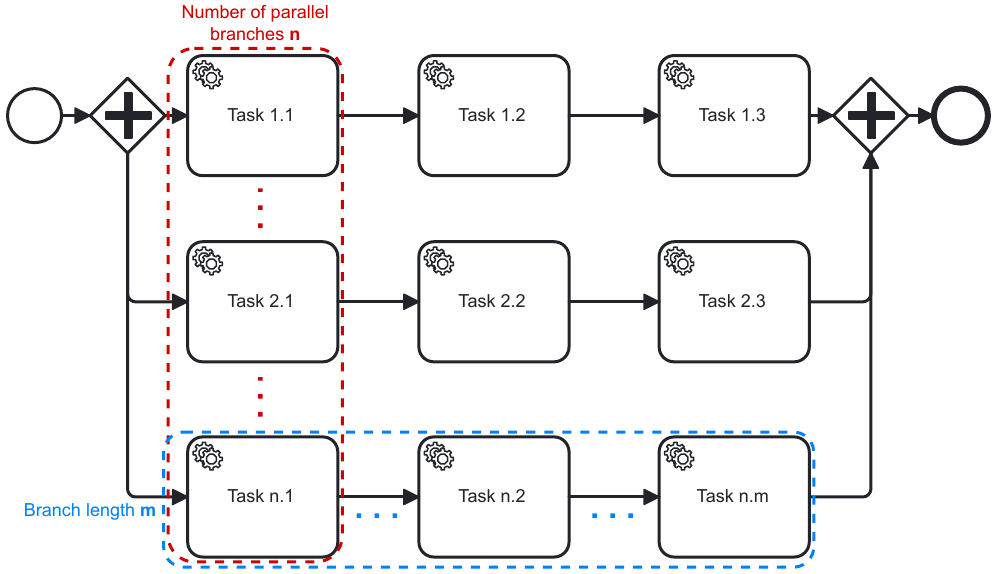}
	\caption{Models with a growing number of parallel branches and branch length}
	\label{fig:parallel-branches-models}
\end{figure}

We generated a data set of models~\cite{timkrauterBPM2024Artifacts2024} with a growing number of parallel branches with increasing length, similar to~\cite{corradiniFormalApproachAnalysis2021}.
\autoref{fig:parallel-branches-models} depicts how a model with \textit{n} parallel branches with length \textit{m} is generated.

\autoref{tab:parallel-branches-benchmark} shows the average runtime of our tool when checking soundness and safeness for BPMN models with a varying number of parallel branches and branch lengths.
Our tool explores the entire state space while simultaneously verifying all properties.

\begin{table}
	\centering
	\caption{Benchmark results of the parallel branches models}
	\label{tab:parallel-branches-benchmark}
	\SetTblrInner{colsep=2pt}
	\begin{tblr}{
			column{1-X} = {c},
			column{Y-Z} = {r},
			hline{1, 2, Z} = {-}{1.2pt, solid}, % Z is the last row/column
			hline{8, 10} = {-}{dashed},
			vline{2-Y} = {2-Z}{solid}, % Y is the second to last row/column
		}
		\textbf{Branches} & \textbf{Branch Length} &\textbf{Runtime} & \textbf{States} \\
		5 & 1 & 1 ms& 35 \\
		10 & 1 & 3 ms& 1.027 \\
		15 & 1 & 161 ms& 32.771 \\
		16 & 1 & 360 ms& 65.539 \\
		17 & 1 & 790 ms& 131.075 \\
		20 & 1 & 8.803 ms& 1.048.579 \\
		5 & 3 & 3 ms& 1.027 \\
		5 & 5 & 14 ms& 7.779 \\
		3 & 10 & 2 ms& 1.334 \\
		3 & 20 & 11 ms& 9.264 \\
	\end{tblr}
\end{table}

The models' state space grows exponentially, leading to the same growth in runtime.
In our case, the number of states is given by $(m+1)^n + 3$ due to our pragmatic encoding (see \autoref{sec:impl}), clearly showing state space explosion.
Our analysis is not instantaneous anymore when approaching 17 parallel branches of length 1 (see \autoref{tab:parallel-branches-benchmark}).
However, analysis is still instantaneous for more reasonable models with five parallel branches of length 5 or 3 branches of length 10-20.
\cite{corradiniFormalApproachAnalysis2021} report 2-3s of runtime for most soundness properties and 30 s for no dead activities for the model with five parallel branches of length 1, which took 1 ms in our tool.
The difference is due to faster hardware and implementation differences, such as our pragmatic encoding leading to smaller state spaces, see \autoref{sec:impl}.
% https://cpu.userbenchmark.com/Compare/Intel-Core-i7-6700HQ-vs-AMD-Ryzen-7-7700X/m34954vs4131

% Comparison to BProve

We believe, as also reported in ~\cite{fahlandAnalysisDemandInstantaneous2011}, that models with a high degree of parallelism are uncommon.
For example, only 3\% of the industrial models in~\cite{fahlandAnalysisDemandInstantaneous2011} have more than 1000 states.
If complex models are more common than anticipated, one can implement \textit{partial order reduction}~\cite{clarkeHandbookModelChecking2018}, which has shown great results for the Petri Net model checker LoLA in~\cite{fahlandAnalysisDemandInstantaneous2011}.
Due to partial order reduction, they reduced the analysis time and checked previously intractable models.
One can implement partial order reduction similarly in our tool since the semantics of BPMN and Petri Nets have many similarities.
However, translating BPMN to Petri nets is not always possible. % ~\cite{favreDifficultyReplacingInclusive2012}.

\subpart{Realistic Models}
We apply our tool to eight realistic models, where three models (e001, e002, e020) are taken from~\cite{houhouFirstOrderLogicVerification2022}, and the remaining five models are part of the Camunda BPMN for research repository\footnote{\url{https://github.com/camunda/bpmn-for-research} \label{footnote:camundaResearch}}.
We had to slightly adapt the Camunda models such that they can be executed standalone, which is described in our artifacts~\cite{timkrauterBPM2024Artifacts2024}.

\autoref{tab:realistic-models-benchmark} shows the average runtime of soundness and safeness checking using our tool and the number of states for each model.
The results show that our tool can check soundness and safeness \textit{instantaneously} for the given realistic models.

\begin{table}
	\centering
	\caption{Benchmark results of the realistic BPMN models}
	\label{tab:realistic-models-benchmark}
	\SetTblrInner{colsep=2pt}
	\begin{tblr}{
			column{2-X, Z} = {c},
			column{X-Y} = {r},
			hline{1, 2, Z} = {-}{1.2pt, solid}, % Z is the last row/column
			hline{5} = {-}{dashed},
			vline{2-Y} = {2-Z}{solid}, % Y is the second to last row/column
		}
		\textbf{Model name} &\textbf{Runtime} & \textbf{States} & \textbf{Violated Properties} \\
		e001~\cite{houhouFirstOrderLogicVerification2022} & 1 ms & 39 & - \\
		e002~\cite{houhouFirstOrderLogicVerification2022} & 1 ms & 39 & - \\
		e020~\cite{houhouFirstOrderLogicVerification2022} & 10 ms & 5356 & - \\
		credit-scoring-async\footref{footnote:camundaResearch} & 1 ms & 60 & - \\
		credit-scoring-sync\footref{footnote:camundaResearch} & 1 ms & 140 & Option To Complete \\
		dispatch-of-goods\footref{footnote:camundaResearch} & 1 ms & 103 & Safeness, Proper Completion\\
		recourse\footref{footnote:camundaResearch} & 1 ms & 77 & - \\
		self-service-restaurant\footref{footnote:camundaResearch} & 1 ms & 190 & - \\
	\end{tblr}
\end{table}

Some models violate soundness properties.
\textit{Safeness} and \textit{Proper Completion} are violated in \textsf{dispatch-of-goods} due to a parallelization which is not synchronized later.
How to automatically fix violated properties is discussed in \autoref{subsec:fixable-checking}.
For example, the violations in \textsf{dispatch-of-goods} are automatically resolvable.

Furthermore, our tool takes 1-10ms for e001, e002, and e020 while~\cite{houhouFirstOrderLogicVerification2022} and~\cite{krauterFormalizationAnalysisBPMN2023} report 3.66-10.26s and 1-1.75s.
The benchmarks in~\cite{krauterFormalizationAnalysisBPMN2023} were run on the same hardware, while the machine used in~\cite{houhouFirstOrderLogicVerification2022} was slightly less powerful.

\subsection{Comprehensible Soundness Checking}

The first step to fixing a soundness violation is understanding the problem.
Thus, a tool must present soundness violations clearly and provide the necessary details to the modeler.
We aim to make soundness checking comprehensible by providing textual feedback and utilizing the BPMN model's graphical structure.

\begin{figure}[ht]
	\centering
	\includegraphics[width=0.9\textwidth]{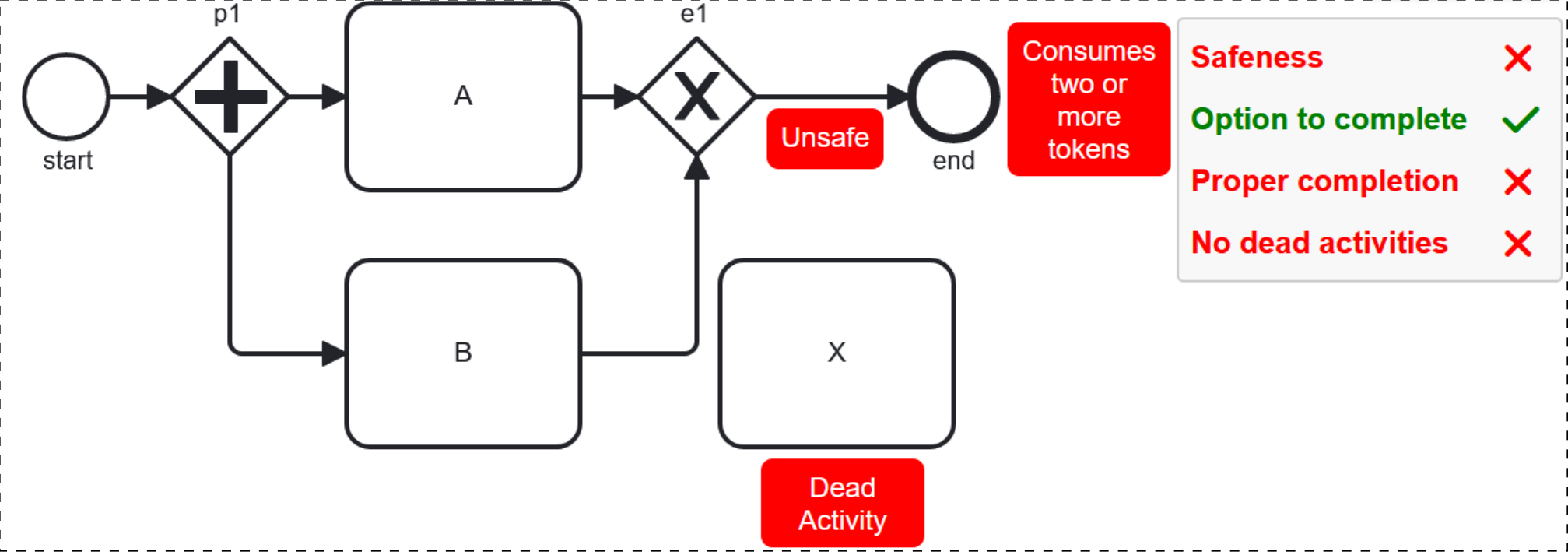}
	\caption{\textit{Soundness violations} in the analysis tool}
	\label{fig:violations}
\end{figure}

% Identify problematic elements using overlays.
We highlight the problematic elements in the BPMN model that cause soundness violations.
\autoref{fig:violations} depicts how we highlight problematic elements using red overlays.
In addition, there is a summary panel in the top-right corner.

% Could be merged with Figure 7 if we need space. Then, it also reduces unnecessary space between elements.

Using the overlays, one can immediately see which elements cause soundness violations, which helps to pinpoint the root problems in the BPMN models.
For \textit{Safeness}, we highlight sequence flows that can become \textit{unsafe}, i.e., two tokens can be located at the sequence flow.
For \textit{Proper Completion}, we identify the end events that can consume more than one token, and for \textit{No Dead Activities}, we highlight the dead activities.
In contrast, we cannot highlight elements for \textit{Option To Complete} since it means the process execution must not always terminate, which cannot easily be attributed to single BPMN elements.

However, just by looking at the BPMN model, it can still be hard to understand soundness violations due to the interplay of the BPMN element's execution semantics.
In the BPMN specification, execution semantics are described using the concept of moving \textit{tokens}.
% Universally used token concept
Tokens are used universally in the industry and research to depict process execution information~\cite{camundaservicesgmbhBpmnjsTokenSimulation2024,corradiniFormalApproachAnalysis2021,corradiniFormalisingAnimatingMultiple2022,houhouFirstOrderLogicVerification2022,krauterFormalizationAnalysisBPMN2023,krauterHigherorderTransformationApproach2024}.

\begin{figure}[ht]
	\centering
	\includegraphics[width=0.9\textwidth]{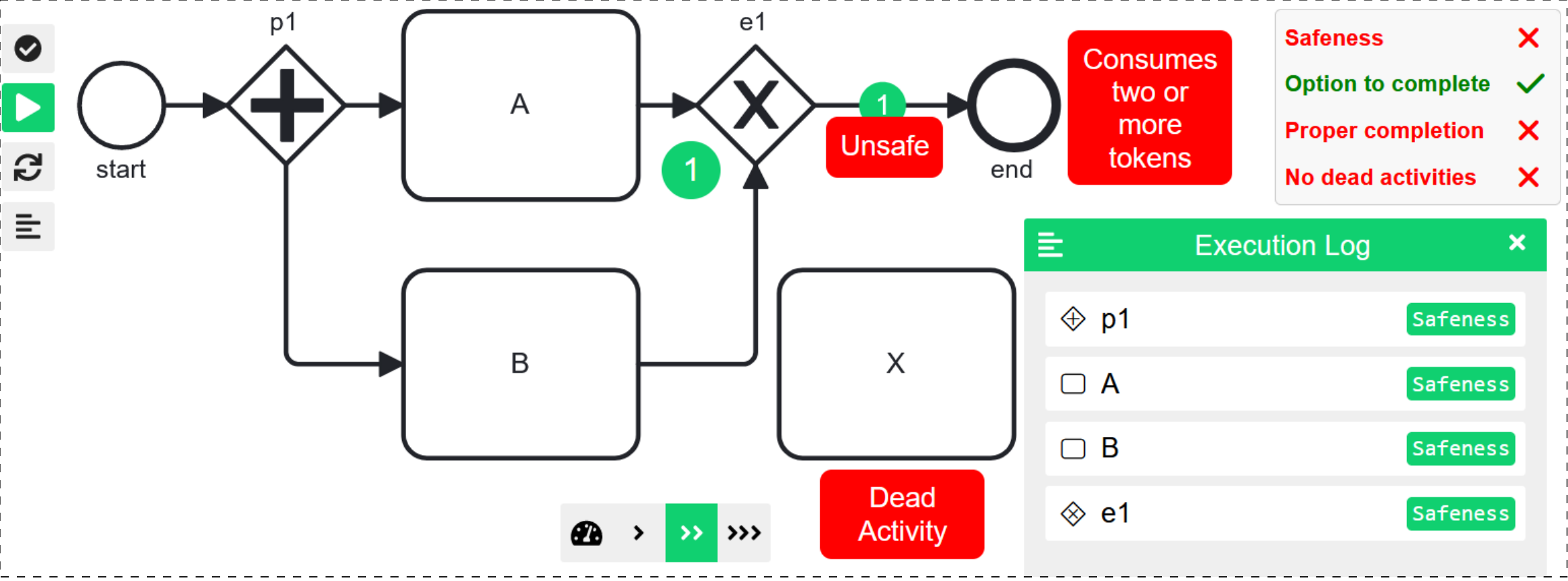}
	\caption{Interactive \textit{counterexample visualization} in the analysis tool}
	\label{fig:counterexample}
\end{figure}

We use tokens in our tool to \textit{interactively} visualize the counterexamples, i.e., violation witnesses of our soundness properties.
Our soundness checker provides counterexamples for all properties except \textit{No Dead Activities}.
Then, we visualize these counterexamples directly in the BPMN editor by showing how tokens move from the process start to a state that violates the given soundness property.
\autoref{fig:counterexample} shows a screenshot visualizing the Safeness counterexample for the same BPMN model as shown in \autoref{fig:violations}.

% A bit wasteful regarding space due to the execution log.

% Describe the figure.
In \autoref{fig:counterexample}, the visualization has been \textit{paused} just before the \textit{unsafe} state was reached.
One token is already located at the sequence flow, which is marked as unsafe, while a second token is currently waiting at the exclusive gateway \textsf{e1}.
The visualization can be resumed or restarted using the play and restart button on the left side.
When resumed, the gateway \textsf{e1} will execute, resulting in two tokens at the subsequent sequence flow, i.e., an unsafe execution state.
In addition, one can control the visualization speed using the bottom buttons next to the speedometer.

% Execution log
Furthermore, our tool shows an \textit{execution log}, which states the history of executing BPMN elements.
In \autoref{fig:counterexample} the parallel gateway \textsf{p1}, the activities \textsf{A} and \textsf{B}, as well as the exclusive gateway \textsf{e1} have each been executed once before the pause.
This is useful since an unsuspected execution order is often the culprit for property violations.

% Conclusion for comprehensible
In summary, we aim to make soundness checking \textit{comprehensible} even for users unaccustomed to the BPMN execution semantics.
First, we directly highlight problematic elements in the model.
Second, we provide an \textit{interactive} visualization of counterexamples using tokens with the ability to pause the visualization, control the visualization speed, and show an execution log.

\subsection{Fixable Soundness Checking} \label{subsec:fixable-checking}

\begin{figure}[ht]
	\centering
	\includegraphics[width=0.5\textwidth]{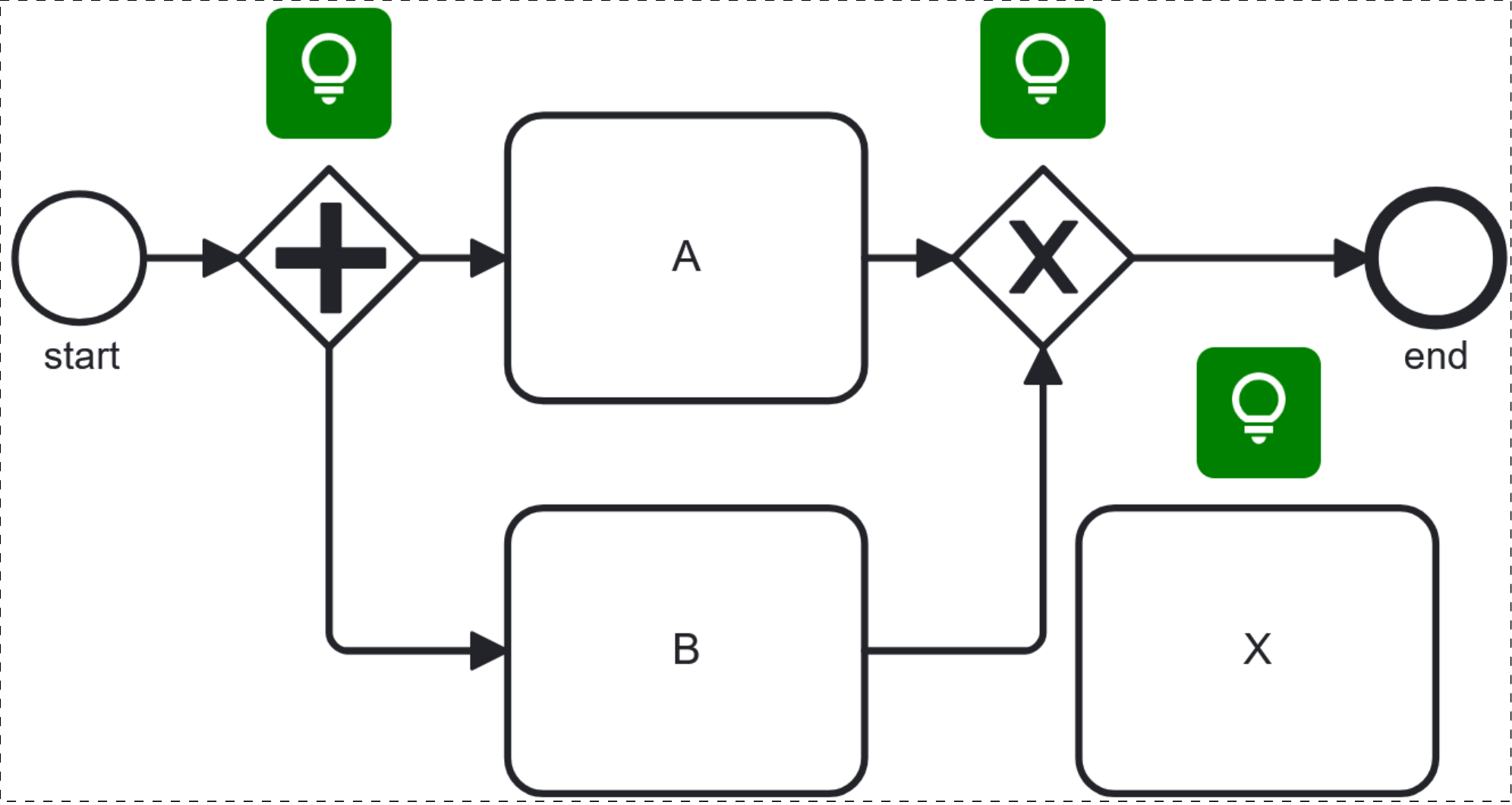}
	\caption{\textit{Quick fixes} in the analysis tool}
	\label{fig:quick-fixes}
\end{figure}

If possible, we provide an automatic fix similar to \textit{quick fixes} in Integrated Development Environments (IDEs) for detected violations.
Modelers can then select these quick fixes to restore soundness.
\autoref{fig:quick-fixes} shows a screenshot of our tool, where quick fixes are depicted as green overlays containing a light bulb icon.
This icon is typically used to indicate quick fixes in IDEs.
Our tool simultaneously shows soundness violations (\autoref{fig:violations}) and quick fixes (\autoref{fig:quick-fixes}).

The user can apply a quick fix by clicking on a green overlay and instantly see the changes in soundness and safeness.
If unhappy with the result, a user can undo all changes since each quick fix is entirely revertible due to the command pattern.
A user might not like a quick fix if it not only fixes a specific property but also has unintended side effects.
For example, it might invalidate a different soundness property.

In the following sections, we describe the implemented resolution strategies for the different soundness properties.
However, we do not expect these strategies to cover all possible resolutions and fit all BPMN models.
Thus, our tool is extensible, so one can provide custom resolution strategies.
This is possible due to its \textit{modular} design compatible with standard mechanisms of the \textit{bpmn.io} ecosystem, which \textit{decouples} the soundness analysis from reporting violations, visualizing counterexamples, and providing quick fixes.

\vspace{-1em}
\subsubsection{Safeness}
The soundness checker will provide a counterexample and identify the unsafe sequence flows.
We use this information together with the structure of the BPMN model to find resolutions for \textit{Safeness} violations.
Possible reasons for \textit{Safeness} violations which we address are:

\begin{enumerate}
	\item An exclusive gateway with multiple incoming sequence flows might be executed more than once, leading to multiple tokens at its outgoing sequence flows.
	\item Similarly, implicitly encoded exclusive gateways exist, for example, if an activity has multiple incoming sequence flows.
	Implicitly encoding exclusive gateways is allowed but violates best practices~\cite{camundaservicesgmbhBpmnlint2024}.
	Similar to \textbf{(1)}, this leads to unsafe sequence flows if the activity is executed more than once.
\end{enumerate}

\textbf{Resolutions for (1):} A straightforward solution is to change the exclusive gateway to a parallel gateway.
Since the gateway was activated multiple times, it indicates that it perhaps should have been a parallel gateway or that there was an unintended parallelization before.
Thus, we can analyze the BPMN model and try to find the parallelization that causes the \textit{Safeness} violation.
If we find a parallel gateway, another solution is to change this parallel gateway to an exclusive one.
This leads to two possible solutions with the overarching goal of matching gateways.

The parallelization can also be implicitly encoded, for example, using an activity with multiple outgoing sequence flows.
This does not comply with best practices~\cite{camundaservicesgmbhBpmnlint2024} but is allowed.
In this case, we can add an explicit exclusive gateway to eliminate the implicit parallelization and achieve matching gateways.

\begin{figure}[ht]
	\centering
	\includegraphics[width=0.8\textwidth]{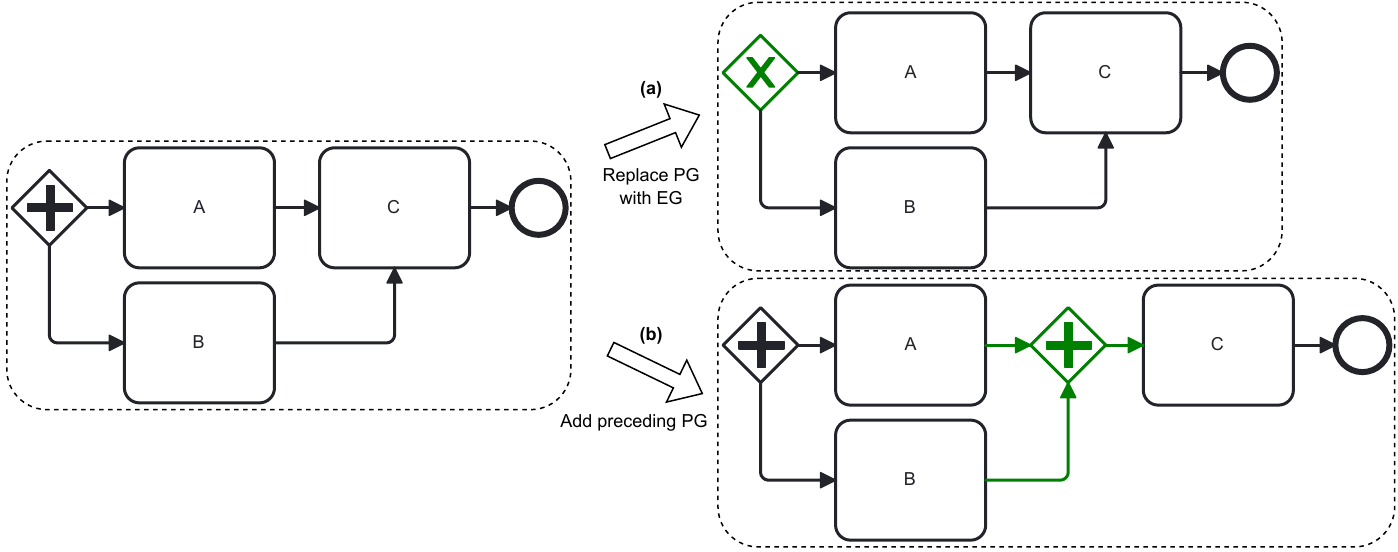}
	\caption{Example quick fixes for \textit{Safeness}}
	\label{fig:safeness}
\end{figure}

\textbf{Resolutions for (2):} Similarly to \textbf{(1)}, the goal of each quick fix is to obtain matching gateways.
This is not always possible since BPMN models must not be well-structured.
Thus, one solution is to find the parallelization that causes the \textit{Safeness} violation and change it to an exclusive gateway, as described in \textbf{(1)}.
Quick fix \textbf{(a)} in \autoref{fig:safeness} shows this solution, where a parallel gateway is changed to an exclusive one.
We color the changes and additions in green to highlight the effect of quick fixes.

Another solution is to remove the implicit exclusive gateway and add a parallel gateway instead, see quick fix \textbf{(b)} in \autoref{fig:safeness}.
The quick fix moves elements automatically to make space to insert the parallel gateway and then reconnects and adds sequence flows accordingly.
Even though these are multiple individual operations, we ensured that an undo operation would revert the entire quick fix.
All the implemented quick fixes can be reverted using one undo operation.

Our demo application in the artifacts~\cite{timkrauterBPM2024Artifacts2024} contains examples of the Safeness quick fixes discussed here and all upcoming soundness properties.

\subpart{Proper Completion}
For \textit{Proper Completion}, our soundness checker will provide a counterexample and identify the problematic end events that consume more than one token.
Possible reasons for \textit{Proper Completion} violations which we address are:

\begin{enumerate}
	\item End event with multiple incoming sequence flows can be executed twice or more.
	This could be due to a parallelization that is never synchronized.
	\item If there is only one incoming sequence flow, then this flow must be unsafe, i.e., hold more than one token in a possible execution.
\end{enumerate}

\begin{figure}[ht]
	\centering
	\includegraphics[width=0.5\textwidth]{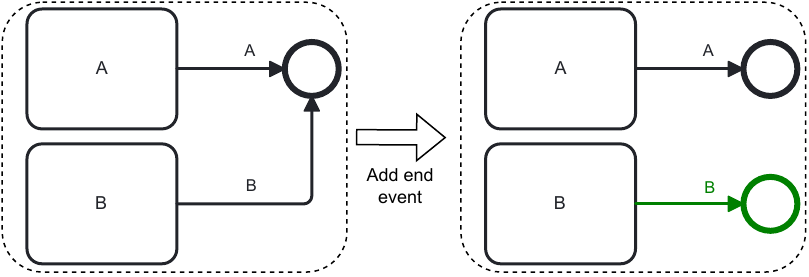}
	\caption{Example quick fix for \textit{Proper Completion}}
	\label{fig:properCompletion}
\end{figure}

\textbf{Resolution for (1):} If multiple incoming sequence flows are the cause, we can add additional end events to match the number of sequence flows.
\autoref{fig:properCompletion} shows an example of applying this quick fix.
We ensure an undo operation would revert the quick fix using the command pattern.

\textbf{Resolution for (2):} If a problematic end event only has one incoming sequence flow, it must be unsafe.
Thus, other \textit{Safeness} quick fixes can apply, which will also resolve \textit{Proper Completion}.

\subpart{Option to Complete}
Violations of \textit{Option to Complete} can have multiple reasons.

\begin{enumerate}
	\item A parallel gateway that synchronizes multiple incoming sequence flows but never executes leads to a violation.
	\item An event that is never triggered but relied upon leads to a violation.
\end{enumerate}

To know the reason for a given violation, we analyze the counterexample provided by the soundness checker.
The counterexample provides a trace that leads to a state in which the process cannot complete.
By analyzing the last state in this trace, i.e., the state in which execution cannot continue, we can determine which element is the cause.
Thus, we can provide quick fixes for the possible reasons.

\begin{figure}[ht]
	\centering
	\includegraphics[width=0.8\textwidth]{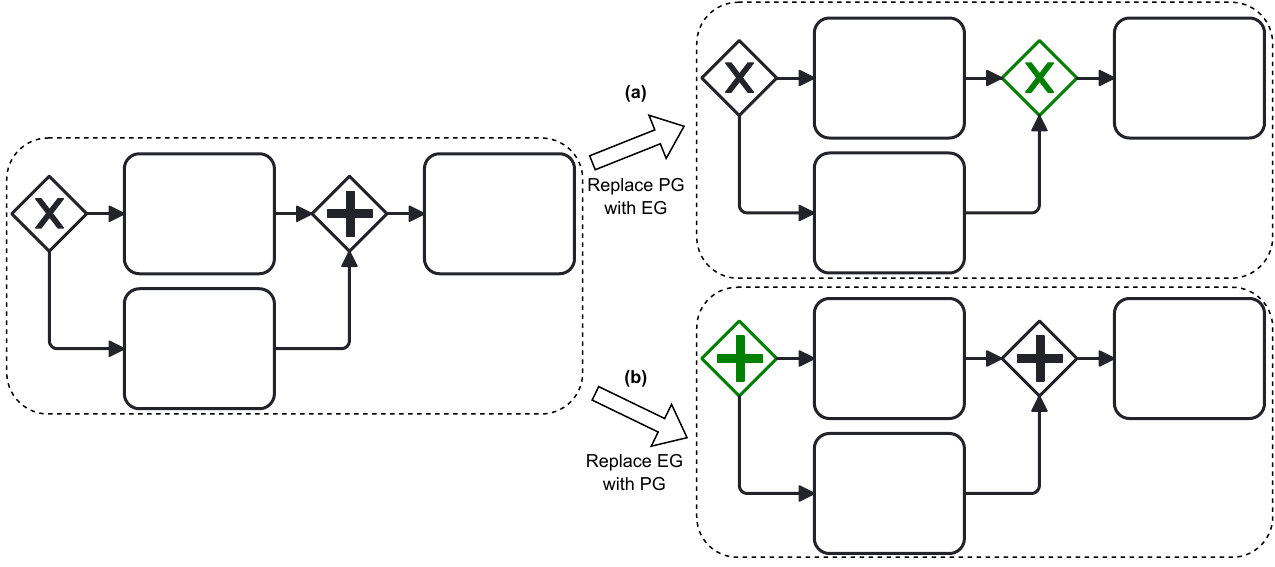}
	\caption{Example quick fixes for \textit{Option To Complete}}
	\label{fig:optionToComplete}
\end{figure}

\textbf{Resolutions for (1):} A straightforward way to fix sequence flow not continuing past a parallel gateway is to change it to an exclusive gateway.
Similar to the Safeness quick fixes, we obtain matching gateways.
Exclusive gateways do not synchronize, and thus, execution can continue.
\autoref{fig:optionToComplete} \textbf{(a)} shows an example of this quick fix.

Another way to fix violations is to find the split in the sequence flow, for example, the exclusive gateway in \autoref{fig:optionToComplete}, and make this split a parallelization, see quick fix \textbf{(b)} in \autoref{fig:optionToComplete}.
We present both possible resolutions to the user, who can choose the appropriate one.
In the example in \autoref{fig:optionToComplete}, it is possible to spot the mismatching gateways.
However, this might not be straightforward in bigger BPMN models with more flow nodes and sequence flows.

\textbf{Resolutions for (2):} We have only implemented quick fixes for untriggered message events since we do not support other intermediate events.
If an untriggered message event has no incoming message flows, we can add a message flow from the closest message throw event or send task to the event.
Currently, we only use spatial proximity to find the right source of the new message flow, but other factors, such as name similarity or reachability, can be considered.
Generalizing this idea to different event types requires adding the missing event trigger.
Analyzing other elements in the model or involving the modeler can help find a potential trigger.

\subpart{No Dead Activities}
A dead activity might have multiple reasons:

\begin{enumerate}
	\item An activity can be disconnected, i.e., it has no incoming sequence flow.
	\item An activity can be a receive task with no incoming message flows.
	\item An activity can also be part of the BPMN model that is not reachable during execution because, for example, a parallel gateway or event preceding the activity cannot be executed.
\end{enumerate}

\textbf{Resolutions for (1)/(2)}: If the activity has no incoming sequence flow/message flow, we can propose adding the missing flow.
A sequence flow is added from the nearest flow node, which is not disconnected or dead, while a message flow is added from the nearest message throw event or send task.

\textbf{Resolutions for (3)}: When the dead activity is connected, this means it is unreachable during execution, and often the process itself cannot terminate, i.e., violates \textit{Option to Complete}.
Thus, other quick fixes can potentially be applied.

\section{Implementation} \label{sec:impl}

In this section, we briefly describe our implementation of the BPMN soundness-checking tool.
The tool's main goal is to be performant, comprehensible, and well-integrated into modeling tools without introducing unnecessary friction.

\subsection{Tool overview}
Our tool is open-source and available as an artifact~\cite{timkrauterBPM2024Artifacts2024}.
The tool architecture is shown in \autoref{fig:overview}, while screenshots of our tool with different features enabled are given in \autoref{fig:violations}, \autoref{fig:counterexample}, and \autoref{fig:quick-fixes}.
The front-end of the tool is built in web technologies using the \textit{bpmn.io} ecosystem, especially the bpmn-js-token-simulation~\cite{camundaservicesgmbhBpmnjsTokenSimulation2024}, while the soundness and safeness checking is implemented in the Rust programming language.

We developed soundness checking in Rust due to its memory efficiency, absence of garbage collection and lightweight hardware abstractions, resulting in an execution performance comparable to C or C++, while retaining memory safety guarantees.
A fast programming language and direct implementation of the BPMN semantics are key in achieving \textit{instantaneous} soundness checking.
We would like to emphasize that the current implementation does not include optimization techniques such as partial order reduction, therefore, we assume that performance may be amplified even further for certain types of models or violations, see, e.g., the parallel branches example.

The tool can be integrated into existing BPMN modeling tools since its model capabilities are invoked through a \textit{web-service} interface or as \textit{command-line application}.
Furthermore, the front-end is \textit{modular} such that one can either fully integrate it, pick only specific features, or extend it, for example, to add custom quick fixes.

\subsection{Soundness checking in Rust}
% State space generation and search at the same time
% \cite{clarkeHandbookModelChecking2018} section 5.2 basic search algorithm. fig1
We implemented a standard breadth-first state space exploration~\cite{clarkeHandbookModelChecking2018}.
While generating the state space, we can check safety properties, such as \textit{Safeness} and \textit{Option To Complete}, \textit{on-the-fly} for each found state~\cite{clarkeHandbookModelChecking2018}.
For example, \autoref{lst:rust} shows how to find unsafe sequence flows in a state to check \textit{Safeness} by inspecting the number of tokens in each state.

\renewcommand\thelstlisting{\arabic{lstlisting}} % Just increasing listing numbers
\begin{lstlisting}[language=Rust, style=boxed, caption={Find the IDs of \textit{unsafe} sequence flows in a given state}, label={lst:rust}]
pub fn find_unsafe_sf_ids(&self) -> Vec<&String> {
    self.snapshots.iter()
        .flat_map(|snapshot| snapshot.tokens.iter())
        .filter(|&(_, token_amount)| *token_amount >= 2)
        .map(|(sf_id, _)| sf_id)
        .collect() }
\end{lstlisting}

\textsf{Self} refers to a \textsf{struct} holding a set of \textsf{snapshots}, i.e., process instances with a map of \textsf{tokens}, that links sequence flow IDs to the number of tokens they are holding.

Similarly, we examine whether a state is stuck for \textit{Option To Complete}.
This means that some process instances are still ongoing, indicated by the presence of tokens, and the state has no outgoing transitions.
In contrast, we check \textit{No Dead Activities} by remembering all executed activities during state space generation and comparing them to all activities in the model. 
Furthermore, \textit{Proper Completion} is checked by remembering executed end events in each state.

\subsection{Pragmatic BPMN Semantics Encoding}
\begin{figure}[ht]
	\centering
	\includegraphics[width=0.65\textwidth]{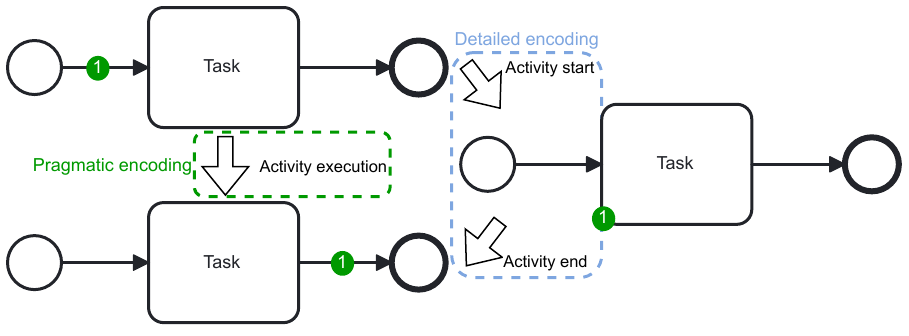}
	\caption{Pragmatic task execution encoding (left) and detailed encoding (right)}
	\label{fig:activityEncoding}
\end{figure}

In addition to selecting an efficient programming language for developing the soundness checker, we also opted for a pragmatic, straightforward encoding of BPMN semantics.
Our encoding contains sufficient details to check soundness and safeness while omitting unnecessary intermediate states, which leads to smaller state spaces.
For example, \autoref{fig:activityEncoding} shows that we do not encode the start and end of a task but rather the execution as a whole.
As discussed earlier, this simple abstraction avoids one additional state and pays off, especially in models with many parallel branches, which can significantly reduce the overall state space.
For example, our tool finds only 2.112 states for the BPMN model \textsf{e020} (see \autoref{subsec:instantaneous}), compared to the 3.558/3.060  in~\cite{houhouFirstOrderLogicVerification2022}/\cite{krauterFormalizationAnalysisBPMN2023}.

Similar minimal encodings can be applied to other BPMN elements, such as message events, to keep the state space small.
Nevertheless, one must always be sure that such optimizations do not compromise the checked properties, similar to when applying partial-order reduction techniques~\cite{clarkeHandbookModelChecking2018}.

Our pragmatic encoding has potential downsides if one wants to check custom properties in the future, such as Activity A and B not executing simultaneously.
Since we are only interested in soundness and safeness, we are consciously trading a smaller state space for the inability to check such custom properties.

\section{Related Work} \label{sec:related-work}

\textbf{BPMN specification coverage:}
Most related work focuses on the depth of BPMN formalization while providing soundness and safeness checking.
Thus, these approaches show how different BPMN elements can be formalized and compare themselves with each other regarding supported elements~\cite{corradiniFormalApproachAnalysis2021,houhouFirstOrderLogicVerification2022,krauterFormalizationAnalysisBPMN2023,krauterHigherorderTransformationApproach2024}.
The supported BPMN elements come close to the capabilities of popular process orchestration platforms.
Our tool supports the same depth of BPMN elements as~\cite{corradiniFormalApproachAnalysis2021}, i.e., the most used gateways, tasks, and events.
A detailed comparison is shown in \autoref{tab:supportedelements}, and we plan to support more elements in the future~\cite{timkrauterBPM2024Artifacts2024}.
We focus on tool performance and capabilities concerning soundness comprehension, soundness resolution, and seamless integration into BPMN modeling tools.

\begin{table}[htbp]
    \caption{BPMN elements supported by different formalizations (based on \cite{vangorpVisualTokenbasedFormalization2013}).}
    \label{tab:supportedelements}
    \begin{tabular}{l c c c c}
    \hline
      BPMN element/feature & Kräuter~\cite{krauterFormalizationAnalysisBPMN2023} &  Corradini~\cite{corradiniFormalApproachAnalysis2021} & Houhou~\cite{houhouFirstOrderLogicVerification2022} & This paper\\
      \hline
      \textit{Instantiation and termination}\\
      Start event instantiation & X & X & X & X \\
      Exclusive event-based gateway & X & & & \\
        \quad instantiation \\
      Parallel event-based gateway & & & & \\
        \quad instantiation \\
      Receive task instantiation & X & & & X \\
      Normal process completion & X & X & X & X \\
      \\
      \textit{Activities} \\
      Activity & X & X & X & X \\
      Subprocess & X & & X & \\
      Ad-hoc subprocesses & & & &\\
      Loop activity & & & & \\
      Multiple instance activity & & & & \\
      Transaction & & & & \\
      \\
      \textit{Gateways} \\
      Parallel gateway & X & X & X & X \\
      Exclusive gateway & X & X & X & X \\
      Inclusive gateway (split) & X & X & X & \\
      Inclusive gateway (merge) & X & & X & \\
      Event-based gateway & X & X & X & X \\
      Complex gateway & & & & \\
      \\
      \textit{Events} \\
      None Events & X & X & X & X \\
      Message events & X & X & X & X \\
      Timer Events & & & X & \\
      Escalation Events & X & & & \\
      Error Events & X & & & \\
      Cancel Events & & & & \\
      Compensation Events & & & & \\
      Conditional Events & & & & \\
      Link Events & X & & & X \\
      Signal Events & X & & & \\
      Multiple Events & & & & \\
      Terminate Events & X & X & X & X \\
      Boundary Events & X & & X & \\
      Event subprocess & X & & & \\
    \end{tabular}
\end{table}

\textbf{Tool performance:}
Comparing tool performance without standard benchmarks and a reproducible environment is challenging.
However, other publications indicate that other tools take several seconds up to half a minute to check single soundness properties~\cite{corradiniFormalApproachAnalysis2021,houhouFirstOrderLogicVerification2022,krauterHigherorderTransformationApproach2024}.
In contrast, our approach instantaneously checks \textit{all} soundness properties and safeness of the same models.
The difference in performance lies probably in our pragmatic BPMN encoding optimized for soundness checking and its direct implementation in a performant programming language rather than in transforming BPMN into general model-checking tools.

\textbf{Tool capabilities:}
Another way to compare the different BPMN formalizations and soundness-checking tools is to investigate their capabilities.
Most tools formalize large parts of the BPMN specification and allow soundness and safeness checking~\cite{corradiniFormalApproachAnalysis2021,houhouFirstOrderLogicVerification2022,krauterFormalizationAnalysisBPMN2023,krauterHigherorderTransformationApproach2024}.
Some tools investigate additional aspects such as the introduction of \textit{data} and \textit{time} during verification~\cite{corradiniFormalisingAnimatingMultiple2022,houhouFirstOrderLogicVerification2022}.
Furthermore, other tools allow specifying and checking \textit{custom temporal logic properties}~\cite{corradiniFormalApproachAnalysis2021,krauterFormalizationAnalysisBPMN2023} and even provide graphical interfaces to ease the specification~\cite{krauterHigherorderTransformationApproach2024}
Moreover, some tools provide interactive BPMN simulation using token-flow animation~\cite{camundaservicesgmbhBpmnjsTokenSimulation2024,corradiniFormalisingAnimatingMultiple2022}, while others \textit{visualize counterexamples} for soundness violations using tokens~\cite{houhouFirstOrderLogicVerification2022}.

Our tool focuses on \textit{instantaneous} soundness and safeness checking and does not support custom properties, data or time.
We do not provide BPMN simulation since other tools already offer this.
However, we use token-flow animation to visualize counterexamples for soundness violations interactively.
This improves comprehension compared to previous static, less interactive visualizations.
In addition, to the best of our knowledge, our tool is the only one that provides \textit{quick fixes}, i.e., automatic resolutions if soundness properties are violated.
To sum up, our tool incorporates and enhances several ideas from the state of the art while adding novel concepts, such as quick fixes.
We advocate for a pragmatic approach, prioritizing performance and understanding above all else to ensure seamless integration into BPMN modeling tools.

\section{Limitations \& Threats to Validity} \label{sec:limitations}
% Limitations
Our tool is a \textit{prototype} that focuses on soundness and safeness and does not check custom properties.
Furthermore, the suggested quick fixes cannot repair all possible violations since the structure of BPMN models allows nearly arbitrary combinations of elements.

Assessing tool performance becomes difficult when there are no standardized benchmarks for comparison.
In this publication, a direct comparison of our tool's performance with related work is not feasible.
This is due to variations in benchmark conditions, such as hardware, operating systems, and employed methodologies.
We claim that our tool can deal with most BPMN models instantaneously, as previously discussed.
It is advisable to consider the results reported by other researchers on the same models only as a general reference.
To solve this problem, we advocate defining a standardized benchmarking process to compare tool performance directly and transparently.

% External validity instantaneous soundness checking:
The main threat to the validity of our instantaneous soundness-checking claim is that we could not test our tool with a large set of models directly from the industry.
To mitigate this threat, we validated our claim against models sourced from existing literature, public repositories, and two artificial datasets.
These datasets, in terms of model size and state space complexity, closely resemble or even surpass industrial models.

\section{Conclusion \& Future work} \label{sec:conclusion}
% Summary and contributions
In this paper, we describe a novel tool that provides instantaneous, comprehensible, and fixable BPMN soundness checking and is integrated into a popular BPMN modeling tool.
We benchmarked our tool against synthetic and realistic BPMN models to demonstrate instantaneous soundness checking.
Our artifacts contain the synthetic data sets of BPMN models and transparently describe our methodology~\cite{timkrauterBPM2024Artifacts2024}.
Our methodology and data sets can be used to better benchmark the performance of soundness-checking tools in the future.

Three main challenges for providing soundness-checking capabilities to non-expert users are identified in~\cite{fahlandAnalysisDemandInstantaneous2011}.
First, a soundness checker must be able to check all or most user-created models, i.e., it must support the most used BPMN elements.
This is not a problem for most tools, including ours, which has similar capabilities to~\cite{corradiniFormalApproachAnalysis2021}, and we plan to increase our tool's BPMN coverage as needed.
Second, soundness checking must be \textit{instantaneous} since long runtimes are unacceptable and often interpreted as tool errors~\cite{fahlandAnalysisDemandInstantaneous2011}.
Third, the biggest challenge for soundness checking is \textit{consumability}, i.e., reporting the found violations in a comprehensible user interface.
Our new tool addresses all these challenges, focusing on instantaneous and comprehensible soundness checking and even providing quick fixes for common soundness violations.
The tool is a BPMN-specific soundness checker written in Rust paired with an intuitive user interface based on the popular \textit{bpmn.io} ecosystem, which allows extending the tool, for example, to provide custom quick fix suggestions.

In future work, we aim to improve our tool by providing more quick fixes, considering advanced BPMN elements such as different events, and ranking quick fixes based on their usefulness.
For example, the impact of quick fixes on soundness properties can be part of the ranking since the resulting model can be checked instantaneously in the background.
Other metrics, such as least change and least surprise from the model repair field, can be used, or one can include previous user behavior.
Finally, we aspire to test our tool in a real-world scenario to gather feedback and measure its impact on productivity.

\bibliographystyle{splncs04}
\bibliography{bib}

\end{document}